\documentclass[onecollarge]{svjour2}       
\smartqed  
\usepackage{graphicx}
\newcommand\gothfamily{\usefont{U}{ygoth}{m}{n}}
\DeclareTextFontCommand{\textgoth}{\gothfamily}
\begin{document}

\title{Conservation laws for a general Lorentz connection
}


\author{Nikodem J. Pop\l awski         
}


\institute{N. J. Pop\l awski \at
              Department of Physics, Indiana University, Swain Hall West 117, 727 East Third Street, Bloomington, Indiana 47405, USA \\
              \email{nipoplaw@indiana.edu}           
}


\maketitle

\begin{abstract}
We derive conservation laws for energy--momentum (canonical and dynamical) and angular momentum for a general Lorentz connection.
\keywords{Conservation law \and Canonical energy--momentum \and Dynamical energy--momentum \and Angular momentum \and Lorentz connection \and Spin density \and Metric compatibility \and Nonmetricity \and Tetrad rotation}
\PACS{04.20.Fy \and 04.50.Kd}
\end{abstract}

\vspace{0.2in}

\section{Introduction}

{\em Gauge} formulation of gravitation~\cite{Uti} attempts to derive a unified picture of known interactions.
This formulation is {\em metric--affine}: both the metric (or tetrad) and affine (or Lorentz) connection are regarded as gravitational potentials~\cite{KS,Hehl,MA}.
Explicit dynamical variables in metric--affine theories can be taken as: metric and symmetric connection (Palatini formulation)~\cite{Pal}, metric and torsion (Einstein--Cartan theory)~\cite{Car}, metric and asymmetric connection~\cite{HK,HLS}, metric, torsion and nonmetricity~\cite{Smal}, tetrad and torsion~\cite{HD}, and tetrad and Lorentz connection (Kibble--Sciama theory)~\cite{KS}.

The principle of general covariance imposes the invariance of the total action under general coordinate transformations. 
Since the metric and tetrad are related by the orthonormality condition, the group of tetrad rotations is the Lorentz group~\cite{L,Lord}.
The local Poincar\'{e} invariance, i.e. the invariance of a Lagrangian density for matter under coordinate transformations and tetrad rotations, leads to identities ({\em conservation laws}) satisfied by matter sources~\cite{HB}.
An energy--momentum conservation (4 equations) results from the coordinate invariance and an angular momentum conservation (6 equations) results from the invariance under tetrad rotations~\cite{L,H}.
The same invariance of the total Lagrangian for matter and gravitational field gives analogous {\em Bianchi identities} satisfied by the field equations.
Since there are $80$ gravitational equations for $16+64=80$ gravitational potentials (tetrad and connection) with $4+6=10$ identities, $80-10=70$ potentials are independent dynamical variables~\cite{L}.

In this paper, which is a sequel of~\cite{Niko}, we present a brief derivation of conservation laws for the canonical and dynamical energy--momentum tensors and angular momentum (spin) density for a {\em general} Lorentz connection.
Such a connection corresponds to an affine connection that is not restricted to be metric compatible and torsionless.
Examples of a physical theory with a general affine connection are Weyl's conformal geometry~\cite{Lord,Weyl} and the generalized Einstein--Maxwell theory with the electromagnetic field tensor represented by the homothetic curvature tensor~\cite{PO}.
We use the notation of~\cite{Niko}.

\section{Infinitesimal coordinate transformations}

Under an infinitesimal coordinate transformation
\begin{equation}
x^\mu\rightarrow x'^\mu=x^\mu+\xi^\mu,
\label{inf}
\end{equation}
where $\xi^\mu$ is an infinitesimal vector, the transformation law for any tensor or tensor density $\Phi$ is given by
\begin{equation}
\delta\Phi=\Phi'(x')-\Phi(x)=\xi^\alpha_{\phantom{\alpha},\beta}C^\beta_\alpha\Phi,
\label{inf1}
\end{equation}
where the constant {\em linear operators} $\hat{C}^\beta_\alpha$ are determined by the {\em covariant derivative} of $\Phi$ with respect to the affine connection~\cite{Lord}:
\begin{equation}
\Phi_{;\mu}=\Phi_{,\mu}+\Gamma^{\,\,\alpha}_{\beta\,\mu} \hat{C}^\beta_\alpha \Phi.
\label{inf2}
\end{equation}
For example, the operator $\hat{C}$ acting on a scalar $\phi$, contravariant vector $V^\nu$, covariant vector $V_\nu$ and scalar density $\mathcal{V}$ of weight $w$ returns, respectively:
\begin{eqnarray}
\hat{C}^\beta_\alpha \phi=0, \\
\hat{C}^\beta_\alpha V^\nu=\delta^\nu_\alpha V^\beta, \\
\hat{C}^\beta_\alpha V_\nu=-\delta^\beta_\nu V_\alpha, \\
\hat{C}^\beta_\alpha \mathcal{V}=-w\delta^\beta_\alpha \mathcal{V}.
\label{operat}
\end{eqnarray}
The transformation law for a Lagrangian density $\textgoth{L}$, which is a scalar density (of weight 1) since $\textgoth{L}d^4x$ is a scalar, is thus
\begin{equation}
\delta \textgoth{L}=-\xi^\mu_{\phantom{\mu},\mu} \textgoth{L}.
\label{Lag1}
\end{equation}

A {\em Lie derivative} with respect to $\xi^\mu$ of a quantity $\Phi$ is defined as
\begin{equation}
\mathcal{L}_\xi \Phi=\bar{\delta}\Phi=\Phi'(x)-\Phi(x)=\delta\Phi-\xi^\nu \Phi_{,\nu}.
\label{Lie1}
\end{equation}
It can be shown that $\bar{\delta}\Phi$, unlike $\delta\Phi$, transforms under general coordinate transformations the same way as $\Phi$.
For example, the Lie derivative of the contravariant metric tensor $g^{\mu\nu}$ is also a tensor:
\begin{equation}
\mathcal{L}_\xi g^{\mu\nu}=\xi^\mu_{\phantom{\mu},\alpha}g^{\alpha\nu}+\xi^\nu_{\phantom{\nu},\alpha}g^{\mu\alpha}-\xi^\alpha g^{\mu\nu}_{\phantom{\mu\nu},\alpha}=2\xi^{(\mu;\nu)}+(4S^{(\mu\nu)}_{\phantom{(\mu\nu)}\alpha}+N^{\mu\nu}_{\phantom{\mu\nu}\alpha})\xi^\alpha.
\label{Lie2}
\end{equation}
A {\em Killing vector} is defined as a vector $\xi^\mu$ that preserves the metric: $\mathcal{L}_\xi g^{\mu\nu}=0$.\footnote{
The invariance of the matter action under a coordinate translation generated by a Killing vector yields the covariant conservation of the corresponding dynamical energy--momentum tensor~\cite{LL}.
}

\section{Canonical energy--momentum tensors}

The change $\delta\textgoth{L}$ of a Lagrangian density for matter under an infinitesimal coordinate transformation~(\ref{inf}) is given by Eq.~(\ref{Lag1}).
If we assume that $\textgoth{L}$ depends, in addition to the coordinates $x^\mu$, on matter fields $\phi$ and their first derivatives $\phi_{,\mu}$ then
\begin{equation}
\delta\textgoth{L}=\frac{\partial\textgoth{L}}{\partial\phi}\delta\phi+\frac{\partial\textgoth{L}}{\partial\phi_{,\mu}}\delta(\phi_{,\mu})+\frac{\bar{\partial}\textgoth{L}}{\partial x^\mu}\xi^\mu,
\label{Lag2}
\end{equation}
where the changes $\delta\phi$ and $\delta(\phi_{,\mu})$ are brought by the transformation~(\ref{inf}) and $\bar{\partial}$ denotes partial differentiation with respect to $x^\mu$ regarding $\phi$ and $\phi_{,\mu}$ as constant.
Using the Lagrange field equations,
\begin{equation}
\frac{\partial\textgoth{L}}{\partial\phi}-\Bigl(\frac{\partial\textgoth{L}}{\partial\phi_{,\mu}}\Bigr)_{,\mu}=0,
\label{LE}
\end{equation}
and the identities $\textgoth{L}_{,\mu}=\frac{\bar{\partial}\textgoth{L}}{\partial x^\mu}+\frac{\partial\textgoth{L}}{\partial\phi}\phi_{,\mu}+\frac{\partial\textgoth{L}}{\partial\phi_{,\nu}}\phi_{,\nu\mu}$ and $\delta(\phi_{,\mu})=(\delta\phi)_{,\mu}-\xi^\nu_{\phantom{\nu},\mu}\phi_{,\nu}$, we bring Eq.~(\ref{Lag2}) to
\begin{equation}
\delta\textgoth{L}=\xi^\mu\textgoth{L}_{,\mu}+\Bigl(\frac{\partial\textgoth{L}}{\partial\phi_{,\mu}}(\delta\phi-\xi^\nu\phi_{,\nu})\Bigr)_{,\mu}.
\label{Lag3}
\end{equation}
Combining Eqs.~(\ref{Lag1}) and~(\ref{Lag3}) gives a conservation law~\cite{Lord}:
\begin{equation}
\textgoth{J}^\mu_{\phantom{\mu},\mu}=\textgoth{J}^\mu_{\phantom{\mu};\mu}-2S_\mu\textgoth{J}^\mu=0,
\label{Noe1}
\end{equation}
with the current:
\begin{equation}
\textgoth{J}^\mu=\xi^\mu\textgoth{L}+\frac{\partial\textgoth{L}}{\partial\phi_{,\mu}}(\delta\phi-\xi^\nu\phi_{,\nu})=\xi^\mu\textgoth{L}+\frac{\partial\textgoth{L}}{\partial\phi_{,\mu}}\bar{\delta}\phi.
\label{Noe2}
\end{equation}
Equations~(\ref{Noe1}) and~(\ref{Noe2}) represent {\em Noether's theorem}: the correspondence between continuous symmetries of a Lagrangian and conservation laws.\footnote{
If $x^\mu$ are Cartesian coordinates then for Lorentz translations, $\xi^\mu=$ const and $\delta\phi=0$, we obtain a conservation of {\em energy--momentum}: $\Theta^\mu_{\nu,\mu}=0$, where $\Theta^\mu_\nu=\frac{\partial\textgoth{L}}{\partial\phi_{,\mu}}\phi_{,\nu}-\delta^\mu_\nu\textgoth{L}$ is the {\em energy--momentum} tensor density.
This conservation also follows from the Lagrange field equations~(\ref{LE}).
For Lorentz rotations, $\xi^\mu=\epsilon^\mu_{\phantom{\mu}\nu}x^\nu$ and $\phi=\frac{1}{2}\epsilon_{\mu\nu}G^{\mu\nu}\phi$, where $G^{\mu\nu}$ are the generators of the Lorentz group, the conservation law~(\ref{Noe1}) yields a conservation of {\em angular momentum}: $(\Lambda^{\phantom{\alpha\beta}\mu}_{\alpha\beta}+\Sigma^{\phantom{\alpha\beta}\mu}_{\alpha\beta})_{,\mu}=0$, where $\Lambda^{\phantom{\alpha\beta}\mu}_{\alpha\beta}=x_\alpha\Theta^\mu_\beta-x_\beta\Theta^\mu_\alpha$ is the {\em orbital angular momentum} density and $\Sigma^{\phantom{\alpha\beta}\mu}_{\alpha\beta}=\frac{\partial\textgoth{L}}{\partial\phi_{,\mu}}G_{\alpha\beta}\phi$ is the {\em spin density}.
}

If we assume that the matter fields $\phi$ are purely tensorial then applying Eqs.~(\ref{inf1}) and~(\ref{inf2}) to the definition of the conserved current~(\ref{Noe2}) gives
\begin{equation}
\textgoth{J}^\mu=\xi^\mu\textgoth{L}+\frac{\partial\textgoth{L}}{\partial\phi_{,\mu}}\Bigl((\xi^\alpha_{\phantom{\alpha};\beta}+2S^\alpha_{\phantom{\alpha}\beta\nu}\xi^\nu)\hat{C}^\beta_\alpha\phi-\xi^\nu\phi_{;\nu}\Bigr).
\label{Noe3}
\end{equation}
In order to obtain a local conservation law that does not contain the vector $\xi^\mu$ we must impose a covariant restriction on this vector at a particular point in spacetime:
\begin{equation}
\xi^\alpha_{\phantom{\alpha};\beta}+2S^\alpha_{\phantom{\alpha}\beta\nu}\xi^\nu=0,
\label{Noe4}
\end{equation}
which brings Eqs.~(\ref{Noe1}) and~(\ref{Noe2}) to
\begin{equation}
\Bigl(\xi^\mu\textgoth{L}-\frac{\partial\textgoth{L}}{\partial\phi_{,\mu}}\xi^\nu\phi_{;\nu}\Bigr)_{;\mu}-2S_\mu\Bigl(\xi^\mu\textgoth{L}-\frac{\partial\textgoth{L}}{\partial\phi_{,\mu}}\xi^\nu\phi_{;\nu}\Bigr)=0.
\label{Noe5}
\end{equation}
Using again Eq.~(\ref{Noe4}) allows to eliminate $\xi^\mu$, leading to
\begin{equation}
\textgoth{L}_{;\nu}-\Bigl(\frac{\partial\textgoth{L}}{\partial\phi_{,\mu}}\phi_{;\nu}\Bigr)_{;\mu}+2S^\rho_{\phantom{\rho}\mu\nu}\frac{\partial\textgoth{L}}{\partial\phi_{,\mu}}\phi_{;\rho}+2S_\mu\frac{\partial\textgoth{L}}{\partial\phi_{,\mu}}\phi_{;\nu}=0,
\label{Noe6}
\end{equation}
which represents a conservation law:
\begin{equation}
\mathcal{H}^\nu_{\mu;\nu}-2S_\nu\mathcal{H}^\nu_\mu+2S^\nu_{\phantom{\nu}\mu\rho}\mathcal{H}^\rho_\nu=0,
\label{Noe7}
\end{equation}
for the {\em canonical energy--momentum} tensor density:\footnote{
The canonical energy--momentum tensor density~(\ref{Noe8}) generalizes the Cartesian energy--momentum tensor density $\Theta^\mu_\nu$ (cf. footnote 2) to a general affine connection, replacing the ordinary derivative $\phi_{,\nu}$ by the covariant derivative $\phi_{;\nu}$.
The corresponding canonical energy--momentum tensor can be symmetrized using the generalized Belinfante--Rosenfeld formula~\cite{HLS,BR}.
}
\begin{equation}
\mathcal{H}^\mu_\nu=\frac{\partial\textgoth{L}}{\partial\phi_{,\mu}}\phi_{;\nu}-\delta^\mu_\nu\textgoth{L}.
\label{Noe8}
\end{equation}

\section{Dynamical energy--momentum tensors}

A {\em dynamical energy--momentum} tensor density in the tetrad formulation of gravity, $\textgoth{T}^a_\mu$, is defined via the variation of a matter Lagrangian density $\textgoth{L}$ with respect to a tetrad:\footnote{
The energy--momentum tensor, canonical or dynamical, is obtained from the corresponding energy--momentum tensor density by dividing the latter by $\textgoth{e}$.
}
\begin{equation}
\delta\textgoth{L}=\textgoth{T}^a_\mu\delta e^\mu_a.
\label{dyn1}
\end{equation}
If $\textgoth{L}$ depends only on tensor matter fields then the tetrad enters $\textgoth{L}$ only where there is the metric tensor, in a combination $g^{\mu\nu}=\eta^{ab}e^\mu_a e^\nu_b$, yielding
\begin{equation}
\delta e^\mu_a=\frac{1}{2}e_{a\nu}\delta g^{\mu\nu}.
\label{dyn2}
\end{equation}
Substituting Eq.~(\ref{dyn2}) to~(\ref{dyn1}) gives the standard general-relativistic form $\delta\textgoth{L}=\frac{1}{2}\textgoth{T}_{\mu\nu}\delta g^{\mu\nu}$, where
\begin{equation}
\textgoth{T}_{\mu\nu}=e_{a\mu}\textgoth{T}^a_{\nu}.
\label{dyn3}
\end{equation}
Therefore, for purely tensorial matter fields, the symmetric part $\textgoth{T}_{(\mu\nu)}$ of the dynamical energy--momentum tensor density in the tetrad formulation coincides with the dynamical energy--momentum tensor density $\mathcal{T}_{\mu\nu}$ in the metric formulation~\cite{Lord,LL}.
If $\textgoth{L}$ depends also on spinor matter fields then Eq.~(\ref{dyn1}) becomes the sum of two parts, tensorial and spinorial:
\begin{equation}
\delta\textgoth{L}=\frac{1}{2}\mathcal{T}_{\mu\nu}\delta g^{\mu\nu}+\textgoth{T}^{a(spin)}_\mu\tilde{\delta} e^\mu_a,
\label{dyn4}
\end{equation}
where $\tilde{\delta} e^\mu_a$ denotes the variation of the tetrad that cannot be related to the variation of the metric tensor.

Let us assume that the matter Lagrangian density $\textgoth{L}$ depends on matter fields $\phi$ (and their first derivatives $\phi_{,\mu}$) that can be expressed in terms of Lorentz and spinor indices only.
Consequently, the tetrad field appears in $\textgoth{L}$ only where there is a derivative of $\phi$, in a covariant combination $e^\mu_a \phi_{|\mu}$.
For example, the Dirac Lagrangian density for a massless particle is $\frac{i}{2}\textgoth{e}(\bar{\psi}\gamma^a e^\mu_a \psi_{|\mu}-e^\mu_a \bar{\psi}_{|\mu}\gamma^a \psi)$ and the Maxwell Lagrangian density $-\frac{1}{4}\sqrt{-\textgoth{g}}F_{\mu\nu}F^{\mu\nu}$, where $F_{\mu\nu}=A_{\nu,\mu}-A_{\mu,\nu}$, can be written as $\frac{1}{2}\textgoth{e}(- e^\mu_a A_{b|\mu} F^{ab}+S^c_{\phantom{c}ab}A_c F^{ab})$.\footnote{
The definition $F_{\mu\nu}=A_{\nu,\mu}-A_{\mu,\nu}$ does not yield $F_{ab}=A_{b,a}-A_{a,b}$.
}
Since $\textgoth{L}=\textgoth{e}L$, where $L$ is a scalar, we obtain\footnote{
For the Maxwell Lagrangian density the variation $\delta L$ due to the variation of the tetrad is: $\delta L=-A_{b|\mu}F^{ab}\delta e^\mu_a$.
}
\begin{equation}
\delta\textgoth{L}=\textgoth{e}\delta L-\textgoth{e}e^a_\mu L\delta e^\mu_a=\textgoth{e}\frac{\partial L}{\partial\phi_{|a}}\phi_{|\mu}\delta e^\mu_a-\textgoth{L}e^a_\mu \delta e^\mu_a=\Bigl(\frac{\partial\textgoth{L}}{\partial\phi_{|a}}\phi_{|\mu}-\textgoth{L}e^a_\mu\Bigr)\delta e^\mu_a.
\label{dyn5}
\end{equation}
Comparing Eq.~(\ref{dyn5}) with~(\ref{dyn1}) shows that the dynamical energy--momentum tensor density $\textgoth{T}^a_\mu$ is a generalized canonical energy--momentum tensor density~\cite{Lord}:
\begin{equation}
\textgoth{T}^a_\mu=\frac{\partial\textgoth{L}}{\partial\phi_{|a}}\phi_{|\mu}-e^a_\mu \textgoth{L},
\label{dyn6}
\end{equation}
or equivalently
\begin{equation}
\textgoth{T}^\mu_\nu=\frac{\partial\textgoth{L}}{\partial\phi_{,\mu}}\phi_{|\nu}-\delta^\mu_\nu \textgoth{L}.
\label{dyn7}
\end{equation}

The canonical energy--momentum tensor density~(\ref{dyn7}) generalizes the density~(\ref{Noe8}), replacing the derivative $\phi_{;\nu}$ with $\phi_{|\nu}$.
The difference between the tensor densities~(\ref{dyn7}) and~(\ref{Noe8}) is
\begin{equation}
\textgoth{T}^\mu_\nu-\mathcal{H}^\mu_\nu=-\frac{\partial\textgoth{L}}{\partial\phi_{,\mu}}\Gamma_\nu\phi,
\label{diff1}
\end{equation}
where the connection $\Gamma_\nu=-\frac{1}{2}\omega_{ab\nu}G^{ab}$ depends on the generators $G^{ab}$ of the representation of the Lorentz group governing a transformation law of $\phi$~\cite{Lord,Niko}.
If the matter fields $\phi$ in the Lagrangian density $\textgoth{L}$ are expressed in terms of coordinate indices only, they must be purely tensorial and the dynamical energy--momentum tensor density~(\ref{dyn6}) corresponds to the dynamical matter density $\mathcal{T}_{\mu\nu}$ in the metric formulation.

\section{Spin density}

For a general connection, not restricted to be metric compatible, the Lorentz connection $\omega^{ab}_{\phantom{ab}\mu}$ is not antisymmetric in the indices $a,b$~\cite{Hehl} and its symmetric part is related to the nonmetricity tensor: $\omega^{(ab)}_{\phantom{(ab)}\mu}=-\frac{1}{2}N^{ab}_{\phantom{ab}\mu}$~\cite{Niko}.
The variation of a matter Lagrangian density $\textgoth{L}$ with respect to a Lorentz connection:
\begin{equation}
\delta\textgoth{L}=\frac{1}{2}\textgoth{S}_{ab}^{\phantom{ab}\mu}\delta \omega^{ab}_{\phantom{ab}\mu},
\label{spin1}
\end{equation}
defines the {\em Lorentz connection-conjugate} density $\textgoth{S}_{ab}^{\phantom{ab}\mu}$.\footnote{
The analogous affine connection-conjugate density $\Pi_{\rho}^{\phantom{\rho}\mu\nu}$, defined via the variation of $\textgoth{L}$ with respect to a general affine connection: $\delta\textgoth{L}=\Pi_{\rho}^{\phantom{\rho}\mu\nu}\delta\Gamma^{\,\,\rho}_{\mu\,\nu}$, is called {\em hypermomentum}~\cite{HLS}.
}
The variation of $\textgoth{L}$ with respect to the antisymmetric part of the Lorentz connection defines the {\em spin density} $\textgoth{M}_{ab}^{\phantom{ab}\mu}$ in the tetrad formulation of gravity:
\begin{equation}
\delta\textgoth{L}=\frac{1}{2}\textgoth{M}_{ab}^{\phantom{ab}\mu}\delta \omega^{[ab]}_{\phantom{[ab]}\mu}.
\label{spin2}
\end{equation}
The variation of $\textgoth{L}$ with respect to the symmetric part of the Lorentz connection, i.e. the nonmetricity tensor, defines the {\em nonmetricity-conjugate} density $\textgoth{N}_{ab}^{\phantom{ab}\mu}$:
\begin{equation}
\delta\textgoth{L}=\frac{1}{2}\textgoth{N}_{ab}^{\phantom{ab}\mu}\delta \omega^{(ab)}_{\phantom{(ab)}\mu}.
\label{spin3}
\end{equation}
The densities $\textgoth{M}_{ab}^{\phantom{ab}\mu}$ and $\textgoth{N}_{ab}^{\phantom{ab}\mu}$ are the symmetric and antisymmetric (in the indices $a,b$) parts of the density $\textgoth{S}_{ab}^{\phantom{ab}\mu}$, respectively:
\begin{equation}
\textgoth{S}_{ab}^{\phantom{ab}\mu}=\textgoth{M}_{ab}^{\phantom{ab}\mu}+\textgoth{N}_{ab}^{\phantom{ab}\mu}.
\label{decomp}
\end{equation}

The Lorentz connection $\omega^{ab}_{\phantom{ab}\mu}$ enters $\textgoth{L}$ only where there is a derivative of $\phi$, in a combination $-\frac{\partial\textgoth{L}}{\partial\phi_{,\mu}}\Gamma_\mu\phi$.
Consequently, the spin density is equal to
\begin{equation}
\textgoth{M}_{ab}^{\phantom{ab}\mu}=\frac{\partial\textgoth{L}}{\partial\phi_{,\mu}}G_{ab}\phi,
\label{spin4}
\end{equation}
and generalizes the spin density $\Sigma^{\phantom{\alpha\beta}\mu}_{\alpha\beta}$ in the Cartesian coordinates (cf. footnote 2).
The difference~(\ref{diff1}) between the tensor densities~(\ref{dyn7}) and~(\ref{Noe8}) is then
\begin{equation}
\textgoth{T}^\mu_\nu-\mathcal{H}^\mu_\nu=\frac{1}{2}\omega^{ab}_{\phantom{ab}\nu}\textgoth{M}_{ab}^{\phantom{ab}\mu}.
\label{diff2}
\end{equation}

\section{Conservation of angular momentum}

The Lorentz group is the group of {\em tetrad rotations}~\cite{Hehl,Lord}.
Since a physical matter Lagrangian density $\textgoth{L}$ is invariant under local, proper Lorentz transformations, it is invariant under tetrad rotations:
\begin{equation}
\delta\textgoth{L}=\frac{\partial\textgoth{L}}{\partial\phi}\delta\phi+\frac{\partial\textgoth{L}}{\partial\phi_{,\mu}}\delta(\phi_{,\mu})+\textgoth{T}^a_\mu\delta e^\mu_a+\frac{1}{2}\textgoth{S}_{ab}^{\phantom{ab}\mu}\delta\omega^{ab}_{\phantom{ab}\mu}=0,
\label{tetrot}
\end{equation}
where the changes $\delta$ correspond to a tetrad rotation.
Under integration of Eq.~(\ref{tetrot}) over spacetime the first two terms vanish because of the field equation for $\phi$~(\ref{LE}):
\begin{equation}
\int\Bigl(\textgoth{T}^a_\mu\delta e^\mu_a+\frac{1}{2}\textgoth{S}_{ab}^{\phantom{ab}\mu}\delta\omega^{ab}_{\phantom{ab}\mu}\Bigr)d^4x=0.
\label{tetrot1}
\end{equation}
For an infinitesimal Lorentz transformation:
\begin{equation}
\Lambda^a_{\phantom{a}b}=\delta^a_b+\epsilon^a_{\phantom{a}b},
\end{equation}
where $\epsilon_{ab}=-\epsilon_{ba}$,
the tetrad $e^a_\mu$ changes according to
\begin{equation}
\delta e^a_\mu=\tilde{e}^a_\mu-e^a_\mu=\Lambda^a_{\phantom{a}b}e^b_\mu-e^a_\mu=\epsilon^a_{\phantom{a}\mu},
\label{tetrot2}
\end{equation}
and the tetrad $e_a^\mu$, because of the identity $\delta(e^a_\mu e_a^\nu)=0$, according to
\begin{equation}
\delta e_a^\mu=-\epsilon_{\phantom{\mu}a}^\mu.
\label{tetrot3}
\end{equation}
The Lorentz connection changes according to
\begin{equation}
\delta\omega^{ab}_{\phantom{ab}\mu}=\delta(e^a_\nu \omega^{\nu b}_{\phantom{\nu b}\mu})=\epsilon^a_{\phantom{a}\nu}\omega^{\nu b}_{\phantom{\nu b}\mu}-e^a_\nu \epsilon^{\nu b}_{\phantom{\nu b};\mu}=\epsilon^a_{\phantom{a}c}\omega^{cb}_{\phantom{cb}\mu}-e^a_\nu \epsilon^{\nu b}_{\phantom{\nu b}|\mu}+\epsilon^a_{\phantom{a}c}\omega^{bc}_{\phantom{bc}\mu}=-\epsilon^{ab}_{\phantom{ab}|\mu}-\epsilon^a_{\phantom{a}c}N^{bc}_{\phantom{bc}\mu}.
\label{tetrot4}
\end{equation}

Substituting Eqs.~(\ref{dyn3}), (\ref{tetrot3}) and~(\ref{tetrot4}) to~(\ref{tetrot1}) gives\footnote{
We also use the partial-integration identity $\int d^4x(\textgoth{V}^\mu)_{|\mu}=2\int d^4x S_\mu\textgoth{V}^\mu$, valid for an arbitrary vector density $\textgoth{V}^\mu$.
}
\begin{eqnarray}
& & 0=-\int\Bigl(\textgoth{T}^a_\mu\epsilon^\mu_{\phantom{\mu}a}+\frac{1}{2}\textgoth{S}_{ab}^{\phantom{ab}\mu}\epsilon^{ab}_{\phantom{ab}|\mu}+\frac{1}{2}\textgoth{S}_{ab}^{\phantom{ab}\mu}\epsilon^a_{\phantom{a}c}N^{bc}_{\phantom{bc}\mu}\Bigr)d^4x \nonumber \\
& & =\int\Bigl(\textgoth{T}_{\mu\nu}\epsilon^{\mu\nu}-\frac{1}{2}\textgoth{S}_{\mu\nu}^{\phantom{\mu\nu}\rho}\epsilon^{\mu\nu}_{\phantom{\mu\nu}|\rho}-\frac{1}{2}\textgoth{S}_{\mu\rho}^{\phantom{\mu\rho}\sigma}\epsilon^{\mu\nu}N^\rho_{\phantom{\rho}\nu\sigma}\Bigr)d^4x \nonumber \\
& & =\int\Bigl(\textgoth{T}_{[\mu\nu]}-S_\rho\textgoth{S}_{[\mu\nu]}^{\phantom{[\mu\nu]}\rho}+\frac{1}{2}\textgoth{S}_{[\mu\nu]\phantom{\rho};\rho}^{\phantom{[\mu\nu]}\rho}+\frac{1}{2}N_{[\mu}^{\phantom{[\mu}\rho\sigma}\textgoth{S}_{\nu]\rho\sigma}\Bigr)\epsilon^{\mu\nu}d^4x.
\label{tetrot5}
\end{eqnarray}
Since the infinitesimal Lorentz rotation $\epsilon^{\mu\nu}$ is arbitrary, we obtain a generalized {\em conservation law for angular momentum} (spin density):\footnote{
If we use the affine connection $\Gamma^{\,\,\rho}_{\mu\,\nu}$, which is invariant under tetrad rotations, instead of the Lorentz connection $\omega^{ab}_{\phantom{ab}\mu}$ as a variable in a Lagrangian density $\textgoth{L}$ then we must replace the term with $\delta\omega^{ab}_{\phantom{ab}\mu}$ in Eq.~(\ref{tetrot}) by a term with $\delta(e^\mu_{a,\nu})$.
The resulting equation is equivalent to Eq.~(\ref{AM}), cf.~\cite{PO}.
}
\begin{equation}
\textgoth{M}_{\mu\nu\phantom{\rho};\rho}^{\phantom{\mu\nu}\rho}=-\textgoth{T}_{\mu\nu}+\textgoth{T}_{\nu\mu}+2S_\rho\textgoth{M}_{\mu\nu}^{\phantom{\mu\nu}\rho}-N_{[\mu}^{\phantom{[\mu}\rho\sigma}\textgoth{M}_{\nu]\rho\sigma}-N_{[\mu}^{\phantom{[\mu}\rho\sigma}\textgoth{N}_{\nu]\rho\sigma}.
\label{AM}
\end{equation}
Eq.~(\ref{AM}) corresponds to the conservation law for the hypermomentum density in the orthonormal gauge~\cite{L}.

\section{Conservation of energy--momentum}

A matter Lagrangian density $\textgoth{L}$ is also invariant under infinitesimal translations of the coordinate system~(\ref{inf}).
The corresponding changes of the tetrad and Lorentz connection are given by Lie derivatives:
\begin{eqnarray}
& & \bar{\delta}e^\mu_a=\mathcal{L}_\xi e^\mu_a=\xi^\mu_{\phantom{\mu},\nu}e^\nu_a-\xi^\nu e^\mu_{a,\nu}, \\
\label{Lietet}
& & \bar{\delta}\omega^{ab}_{\phantom{ab}\mu}=\mathcal{L}_\xi \omega^{ab}_{\phantom{ab}\mu}=-\xi^\nu_{\phantom{\nu},\mu}\omega^{ab}_{\phantom{ab}\nu}-\xi^\nu \omega^{ab}_{\phantom{ab}\mu,\nu}.
\label{Liecon}
\end{eqnarray}
Eq.~(\ref{tetrot1}) becomes now
\begin{equation}
\int\Bigl(\textgoth{T}^a_\mu\bar{\delta} e^\mu_a+\frac{1}{2}\textgoth{S}_{ab}^{\phantom{ab}\mu}\bar{\delta}\omega^{ab}_{\phantom{ab}\mu}\Bigr)d^4x=0,
\label{trans1}
\end{equation}
and holds for an arbitrary vector $\xi^\mu$:
\begin{eqnarray}
& & 0=\int\Bigl(\textgoth{T}^a_\mu \xi^\mu_{\phantom{\mu},\nu}e^\nu_a-\textgoth{T}^a_\mu \xi^\nu e^\mu_{a,\nu}-\frac{1}{2}\textgoth{S}_{ab}^{\phantom{ab}\mu} \xi^\nu_{\phantom{\nu},\mu}\omega^{ab}_{\phantom{ab}\nu}-\frac{1}{2}\textgoth{S}_{ab}^{\phantom{ab}\mu} \xi^\nu \omega^{ab}_{\phantom{ab}\mu,\nu}\Bigr)d^4x \nonumber \\
& & =\int\Bigl(-\textgoth{T}^\nu_{\phantom{\nu}\mu,\nu}-\textgoth{T}^a_\nu e^\nu_{a,\mu}+\frac{1}{2}(\textgoth{S}_{ab}^{\phantom{ab}\nu}\omega^{ab}_{\phantom{ab}\mu})_{,\nu}-\frac{1}{2}\textgoth{S}_{ab}^{\phantom{ab}\nu}\omega^{ab}_{\phantom{ab}\nu,\mu}\Bigr)\xi^\mu d^4x.
\label{trans2}
\end{eqnarray}
Consequently we can write
\begin{eqnarray}
& & 0=\textgoth{S}_{ab\phantom{\nu},\nu}^{\phantom{ab}\nu}\omega^{ab}_{\phantom{ab}\mu}+\textgoth{S}_{ab}^{\phantom{ab}\nu}(\omega^{ab}_{\phantom{ab}\mu,\nu}-\omega^{ab}_{\phantom{ab}\nu,\mu})-2\textgoth{T}^\nu_{\phantom{\nu}\mu,\nu}-2\textgoth{T}^a_\nu e^\nu_{a,\mu} \nonumber \\
& & =(\textgoth{S}_{ab\phantom{\nu}|\nu}^{\phantom{ab}\nu}-2S_\rho \textgoth{S}_{ab}^{\phantom{ab}\rho}+\textgoth{S}_{cb}^{\phantom{cb}\nu}\omega^c_{\phantom{c}a\nu}+\textgoth{S}_{ac}^{\phantom{ac}\nu}\omega^c_{\phantom{c}b\nu})\omega^{ab}_{\phantom{ab}\mu}+\textgoth{S}_{ab}^{\phantom{ab}\nu}(-R^{ab}_{\phantom{ab}\mu\nu}+\omega^a_{\phantom{a}c\mu}\omega^{cb}_{\phantom{cb}\nu}-\omega^a_{\phantom{a}c\nu}\omega^{cb}_{\phantom{cb}\mu}) \nonumber \\
& & -2\textgoth{T}^\nu_{\phantom{\nu}\mu,\nu}-2\textgoth{T}^a_\nu e^\nu_{a,\mu},
\label{trans3}
\end{eqnarray}
which reduces to
\begin{eqnarray}
& & 0=(\textgoth{S}_{ab\phantom{\nu}|\nu}^{\phantom{ab}\nu}-2S_\rho \textgoth{S}_{ab}^{\phantom{ab}\rho}-\textgoth{S}_{ac}^{\phantom{ac}\nu}N_{b\phantom{c}\nu}^{\phantom{b}c})\omega^{ab}_{\phantom{ab}\mu}-R^{ab}_{\phantom{ab}\mu\nu}\textgoth{S}_{ab}^{\phantom{ab}\nu} \nonumber \\
& & -2\textgoth{T}^\nu_{\phantom{\nu}\mu;\nu}+4S_\nu \textgoth{T}^\nu_{\phantom{\nu}\mu}-2\textgoth{T}_{\rho\nu}\omega^{\nu\rho}_{\phantom{\nu\rho}\mu}+4S^{\nu\rho}_{\phantom{\nu\rho}\mu}\textgoth{T}_{\rho\nu} \nonumber \\
& & =(\textgoth{M}_{\alpha\beta\phantom{\nu};\nu}^{\phantom{\alpha\beta}\nu}+\textgoth{N}_{\alpha\beta\phantom{\nu};\nu}^{\phantom{\alpha\beta}\nu}-2S_\rho \textgoth{S}_{\alpha\beta}^{\phantom{\alpha\beta}\rho}-\textgoth{S}_{\alpha\sigma}^{\phantom{\alpha\sigma}\nu}N_{\beta\phantom{\sigma}\nu}^{\phantom{\beta}\sigma})\omega^{\alpha\beta}_{\phantom{\alpha\beta}\mu}-R^{\alpha\beta}_{\phantom{\alpha\beta}\mu\nu}\textgoth{S}_{\alpha\beta}^{\phantom{\alpha\beta}\nu} \nonumber \\
& & -2\textgoth{T}^\nu_{\phantom{\nu}\mu;\nu}+4S_\nu \textgoth{T}^\nu_{\phantom{\nu}\mu}-2\textgoth{T}_{\rho\nu}\omega^{\nu\rho}_{\phantom{\nu\rho}\mu}+4S^{\nu\rho}_{\phantom{\nu\rho}\mu}\textgoth{T}_{\rho\nu}.
\label{trans4}
\end{eqnarray}

Equations~(\ref{decomp}), (\ref{AM}) and the identity\footnote{
This identity follows from the formula for the commutator of the covariant derivatives of the metric tensor.
}
$R^{(\alpha\beta)}_{\phantom{(\alpha\beta)}\mu\nu}=N^{\alpha\beta}_{\phantom{\alpha\beta}[\mu;\nu]}+S^\rho_{\phantom{\rho}\mu\nu}N^{\alpha\beta}_{\phantom{\alpha\beta}\rho}$ bring Eq.~(\ref{trans4}) to the form of a generalized {\em conservation law for energy--momentum}:
\begin{eqnarray}
& & \textgoth{T}^\nu_{\phantom{\nu}\mu;\nu}=2S_\nu \textgoth{T}^\nu_{\phantom{\nu}\mu}-2S^\nu_{\phantom{\nu}\mu\rho}\textgoth{T}^\rho_{\phantom{\rho}\nu}+\frac{1}{2}\textgoth{T}_{\nu\rho}N^{\nu\rho}_{\phantom{\nu\rho}\mu}-\frac{1}{4}\textgoth{N}_{\nu\rho\phantom{\sigma};\sigma}^{\phantom{\nu\rho}\sigma}N^{\nu\rho}_{\phantom{\nu\rho}\mu}+\frac{1}{4}\textgoth{M}_{\alpha\rho\sigma}N_\beta^{\phantom{\beta}\rho\sigma}N^{\alpha\beta}_{\phantom{\alpha\beta}\mu} \nonumber \\
& & +\frac{1}{4}\textgoth{N}_{\alpha\rho\sigma}N_\beta^{\phantom{\beta}\rho\sigma}N^{\alpha\beta}_{\phantom{\alpha\beta}\mu}+\frac{1}{2}\textgoth{N}_{\alpha\beta}^{\phantom{\alpha\beta}\rho}S_\rho N^{\alpha\beta}_{\phantom{\alpha\beta}\mu}-\frac{1}{2}\textgoth{M}_{\alpha\beta}^{\phantom{\alpha\beta}\nu}R^{\alpha\beta}_{\phantom{\alpha\beta}\mu\nu}-\frac{1}{2}\textgoth{N}_{\alpha\beta}^{\phantom{\alpha\beta}\nu}(N^{\alpha\beta}_{\phantom{\alpha\beta}[\mu;\nu]}+S^\rho_{\phantom{\rho}\mu\nu}N^{\alpha\beta}_{\phantom{\alpha\beta}\rho}).
\label{EM}
\end{eqnarray}
Eq.~(\ref{EM}) corresponds to the conservation law for the canonical energy--momentum density in the orthonormal gauge~\cite{L}, cf. also~\cite{PO,S}.

The conservation law~(\ref{EM}) for the dynamical energy--momentum tensor density $\textgoth{T}^\nu_{\phantom{\nu}\mu}$ coincides with the conservation law~(\ref{Noe7}) for the canonical energy--momentum density $\mathcal{H}^\nu_\mu$ if the nonmetricity tensor $N_{\mu\nu\rho}$ and the spin density $\textgoth{M}_{\mu\nu}^{\phantom{\mu\nu}\rho}$ vanish.
In order to derive Eq.~(\ref{Noe7}) we assumed that the matter fields $\phi$ are purely tensorial and only considered a coordinate translation in the variation of $\phi$.
Therefore the spin density does not contribute to the conservation of the tensor density $\mathcal{H}^\nu_\mu$ and does not appear in Eq.~(\ref{Noe7}).
In fact, the difference~(\ref{diff2}) between the tensor densities~(\ref{dyn7}) and~(\ref{Noe8}) is linear in the spin density.
The absence of the nonmetricity tensor in Eq.~(\ref{Noe7}) is related to the fact that we imposed the constraint~(\ref{Noe4}) on the vector $\xi^\mu$ to derive a covariant conservation law independent of $\xi^\mu$, while there is no restriction on this vector in the derivation of the conservation law~(\ref{EM}).
The full conservation law for $\mathcal{H}^\nu_\mu$ (corresponding to unrestricted $\xi^\mu$) can be derived by combining Eqs.~(\ref{diff2}), (\ref{AM}) and~(\ref{EM}).



\end{document}